\documentclass[iop]{emulateapj}

\slugcomment{Accepted to ApJ}

\shorttitle{HD 259440: The Proposed Optical Counterpart of HESS J0632+057}
\shortauthors{Aragona et al.}

\begin{document}


\title{HD 259440: The Proposed Optical Counterpart of the $\gamma$-ray Binary HESS J0632+057}


\author{Christina Aragona, M.\ Virginia McSwain}
\affil{Department of Physics, Lehigh University, 16 Memorial Drive East, Bethlehem, PA 18015}
\email{cha206@lehigh.edu, mcswain@lehigh.edu}

\author{Micha\"el De Becker}
\affil{Institut d'Astrophysique et G\'eophysique, Universit\'e de Li\'ege, B-4000 Li\'ege, Belgium}
\affil{Observatoire de Haute-Provence, F-04870 Saint-Michel l'Observatoire, France}
\email{debecker@astro.ulg.ac.be}

\begin{abstract}
HD 259440 is a B0pe star that was proposed as the optical counterpart to the $\gamma$-ray source HESS J0632+057.  Here we present optical spectra of HD 259440 acquired to investigate the stellar parameters, the properties of the Be star disk, and evidence of binarity in this system.  Emission from the H$\alpha$ line shows evidence of a spiral density wave in the nearly edge-on disk.  We find a best fit stellar effective temperature of 27500$-$30000 K and a log surface gravity of 3.75$-$4.0, although our fits are somewhat ambiguous due to scattered light from the circumstellar disk.  We derive a mass of 13.2$-$19.0 $M_\odot$ and a radius of 6.0$-$9.6 $R_\odot$.  By fitting the spectral energy distribution, we find a distance between 1.1$-$1.7 kpc.  We do not detect any significant radial velocity shifts in our data, ruling out orbital periods shorter than one month.  If HD 259440 is a binary, it is likely a long period ($> 100$ d) system.
 \end{abstract}




\section{Introduction \label{intro}}

HD 259440 (=MCW 443, BD +05 1291), which lies between  the edge of the Monoceros Loop supernova remnant and the star forming region of the Rosette Nebula, is classified as a B0pe star by \citet{morgan} and a BN0 nne star by \citet{turner}.  \citet{gutier} assign it a projected rotational velocity, $V \sin i$, of 430 km s$^{-1}$.  Otherwise, no detailed optical study of this star has been performed.  HD 259440 has been proposed as the optical counterpart to a possible new $\gamma$-ray binary, HESS J0632+057 \citep{aharonian}.  $\gamma$-ray binaries are a subset of high-mass X-ray binaries which also emit radiation in the MeV-TeV range.  The very high energy emission in such objects is expected to originate from an interaction between the massive star's wind and the compact object, although the physical processes that produce this radiation are under debate (see e.g. \citealt{dubus}).  

Observations taken with the HESS telescope array between 2004 March and 2006 March revealed the TeV point source HESS J0632+057 \citep{aharonian}.  \citet{hinton} followed up this detection with a 26 ks observation of the field using \emph{XMM-Newton} on 2007 September 17.  XMMU J063259.3+054801, the brightest source in the field, was positionally coincident with the HESS observation and centered on HD 259440.  The X-ray source showed a gradual decline in flux over the course of the observation.  X-ray observations obtained with \emph{Swift} showed flux variations on a variety of timescales, from several days to a month \citep{falcone}.  While the data did not show a clear periodicity, \citet{falcone} determined that periods of 35$-$40 days or $\geq$ 54 days are possibilities.  HESS J0632+057 has also been associated with the \emph{ROSAT} source 1RXS J063258.3+054857 and the EGRET source 3EG J0634+0521 \citep{aharonian, skilton}.  The source has not yet been detected by \emph{Fermi} \citep{abdo}.  A non-detection by VERITAS indicates a decrease in $\gamma$-ray flux from the time of the original HESS detection as well \citep{acciari}.  Radio observations revealed the source to have a mean flux of 0.3 mJy at 5 GHz, with variations on the timescale of a month \citep{skilton}.  The multiwavelength detections and the flux variations of this system are consistent with the class of $\gamma$-ray binaries. 

Currently, six systems are known or proposed $\gamma$-ray binaries: LS 5039, LS I +61 303 \citep{abdo}, PSR B1259$-$63 \citep{aharonian2005}, Cyg X-1 \citep{albert}, Cyg X-3 \citep{abdoa}, and HESS J0632+057 (see above).  Optical spectroscopy can provide useful information about the stellar properties and orbital parameters of these systems, which aid in the analysis of observations at other wavelengths.  We have undertaken a study of the optical properties of HD 259440 to complement ongoing work in X-ray, $\gamma$-ray, and radio to determine the nature of this system.

In order to determine the properties of HD 259440, we present an extensive collection of red and blue optical spectra.  Section 2 describes our observations.  Section 3 details our search for radial velocity variations and places limits on the orbital period for the system.  We address the circumstellar disk features in Section 4.  In Section 5, we discuss our fitting procedure using model spectra to determine the stellar properties.  Section 6 describes the spectral energy distribution (SED) of the Be star.  We summarize our results in Section 7.


\section{Observations \label{observations}}

We  observed HD 259440 using the Kitt Peak National Observatory (KPNO) Coud\'e Feed (CF) telescope with the F3KB detector between 2008 October 17 and November 21.  We used both blue and red spectral setups each night and generally obtained two spectra of HD 259440 in each.  The blue spectra were taken using grating B in the third order with the 4-96 order sorting filter.  They have a resolving power $R = \lambda/\Delta\lambda \sim 9500$, cover a wavelength range $4130-4570$ \AA, and have exposure times ranging from 20$-$30 minutes giving a signal-to-noise, S/N, of 50--140.  The red spectra were obtained using grating B in the second order and the OG550 order sorting filter.  They have $R \sim 12000$, span a wavelength range of  $6400-7050$ \AA, and have exposure times of 20 minutes giving a S/N of 100$-$250.  ThAr comparison lamp spectra were taken every 2$-$3 hours for wavelength calibrations.  We zero corrected, flatfielded, and wavelength calibrated each spectrum using standard procedures in IRAF\footnote{IRAF is distributed by the National Optical Astronomy Observatory, which is operated by the Association of Universities for Research in Astronomy, Inc., under cooperative agreement with the National Science Foundation.}.  All spectra were interpolated to a common wavelength grid and corrected for heliocentric radial velocity variations.  The blue spectra were written to a common log wavelength scale for radial velocity analysis.  The spectra were rectified in IRAF using emission line free continuum regions. 

Two blue optical spectra were obtained at the KPNO 2.1m telescope using the GoldCam Spectrograph on 2008 December $12-13$.   Grating G47 was used in the second order with a CuSO$_{4}$ blocking filter and a slit width of $1.3"$ to achieve $R \sim 2100-3100$. These observations cover a wavelength range of $3700-4900$ \AA.  The two spectra have exposure times of 15$-$20 minutes, resulting in a S/N of 125$-$150. Comparison spectra were taken using a HeNeAr lamp to wavelength calibrate the data.  The spectra were zero corrected, flatfielded, and wavelength calibrated using the standard procedures in IRAF.  We corrected the spectra for heliocentric radial velocity variations and rectified them in IRAF using emission line free continuum regions.

We also observed HD 259440 seven times, twice in 2009 March and five times in 2009 October, with the SOPHIE cross-dispersed \'echelle spectrograph at the Observatoire Haute-Provence (OHP) 1.93m telescope.  These observations used the 52.65 l~mm$^{-1}$ R2 \'echelle grating to cover 39 orders in the high efficiency mode ($R = 40000$) covering a wavelength range of $3872-6943$ \AA.  The exposure times were between 45$-$90 minutes long, yielding a S/N of 80$-$220.  Additional spectra were obtained from the OHP archive, which were taken between 2007 October and 2008 February.  These spectra had shorter exposure times of 5$-$15 minutes, and S/N as low as 40.  The spectra were corrected for blaze, flatfielded, wavelength calibrated using ThAr lamp spectra taken within 2 hours of each observation, and corrected for heliocentric radial velocity variations.  The observations are summarized in Table \ref{obs}.


\section{Constraints on Orbital Period \label{orbparams}}

We performed a cross-correlation of the \ion{He}{1} $\lambda 4471$ line in our CF data with a mean spectrum from that run to search for radial velocity, $V_R$, shifts.  Due to the low S/N of those spectra, the $V_R$ measurements had a standard deviation, $\sigma_{RV}$, of 20.1 km s$^{-1}$, and no significant shifts in $V_R$ were found.  We also did not detect any gradual trend of increasing or decreasing $V_R$.  The CF dataset rules out orbital periods of $\sim 35$ d or less.

The \ion{He}{1} $\lambda$5876 line in the \'echelle spectra from OHP was also examined for $V_R$ variations, of which the results are presented in Table \ref{vrmes}.   These measurements had a $\sigma_{RV}$ of 6.7 km s$^{-1}$, which is close to the precision that we would expect to achieve due to the natural line width and line shape.  Again, no significant shifts in $V_R$ were detected in this set of observations.  Due to the significant instrumental differences and wavelength coverage between the CF, KPNO 2.1m, and OHP runs, we hesitate to combine the $V_R$ over these combined datasets to investigate $V_R$ shifts.  \citet{crampton} gives a radial velocity measurement of 7.6 km/s; however, the measurement is flagged for poor quality.  Thus we use only the sparsely sampled OHP spectra to constrain the orbital period further.  

We cannot rule out the possibility that HD 259440 is a binary until we determine the probability that the system has an orbit with $V_R$ shifts too small for our observations to detect.  We use a method similar to that outlined in \citet{garmany}.  For a binary system, the mass function of the primary is 
\begin{equation}
f(m)=\frac{M_\star~\sin^{3}i}{q~(1+q)^2}=1.0355 \times 10^{-7}~K_\star^3~P~(1-e^2)^{3/2}
\end{equation}
with 
\begin{equation}
q=\frac{M_\star}{M_X}
\end{equation}
where $M_\star$ is the mass of the optical star in solar units, $M_X$ is the mass of the compact object in solar units, $K_\star$ is the velocity semi-amplitude of the optical star in km s$^{-1}$, $P$ is the orbital period in days, and $e$ is the orbital eccentricity.  Assuming an upper limit for $K_\star$ of $2\sigma_{RV}$, we can find the upper limit for the system inclination for a given value of $P$ at which we would fail to detect $V_R$ variations in our dataset:
\begin{equation}
\sin i \leq 9.392 \times 10^{-3} \sigma_{RV} \left(\frac{P q (1+q)^2 (1-e^2)^{3/2}}{M_\star}\right)^{1/3}\label{eqnsi}.
\end{equation}
If we assume a random distribution of orbital inclinations, the probability that $V_R$ is lower than our detection limit is
\begin{equation}
P(i < i_{up}) = \int_{0}^{i_{up}} \sin i~di~=~1-\cos i_{up}.
\end{equation}

\begin{figure}
\includegraphics[angle=180,scale=0.35]{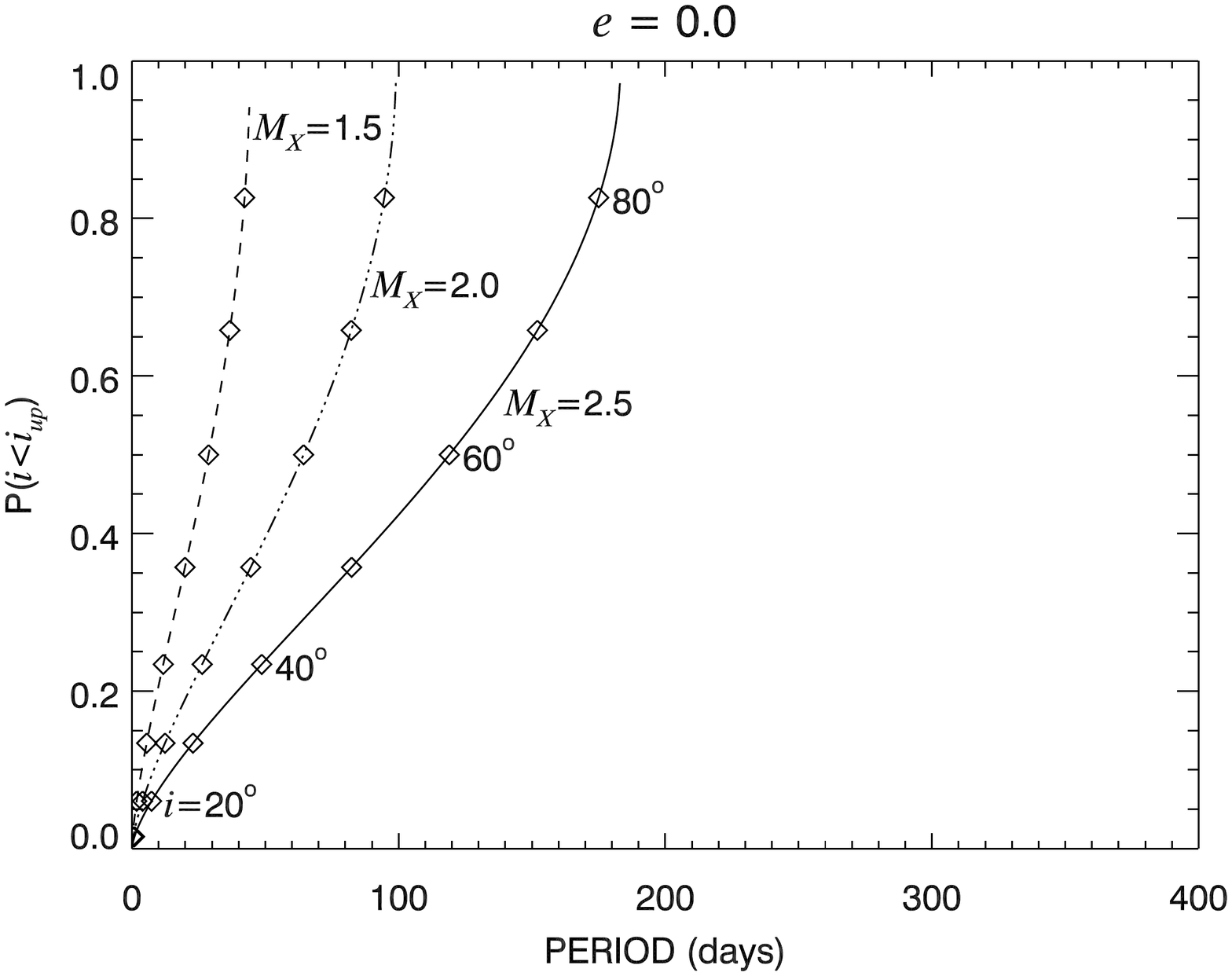} 
\includegraphics[angle=180,scale=0.35]{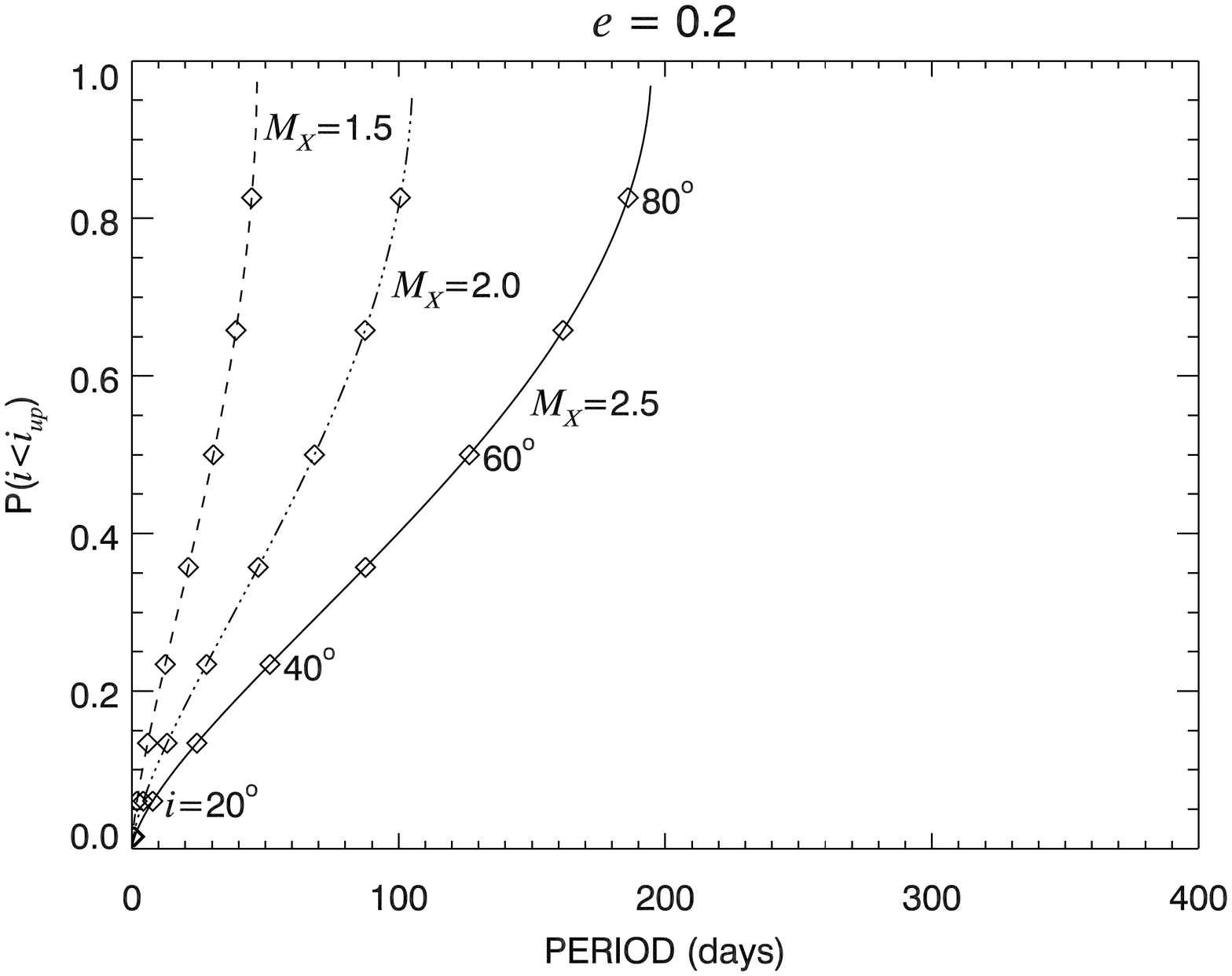}
\includegraphics[angle=180,scale=0.35]{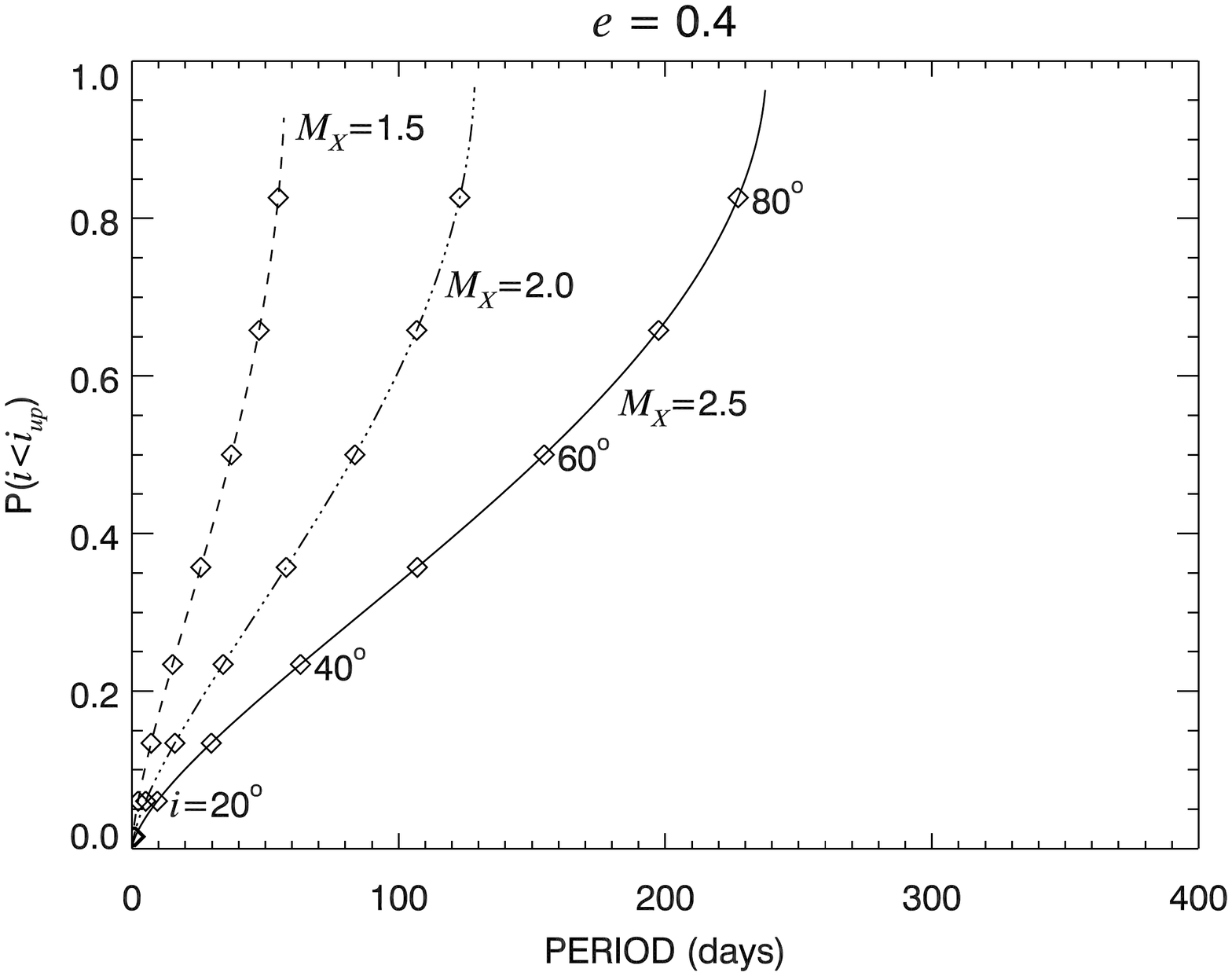}
\includegraphics[angle=180,scale=0.35]{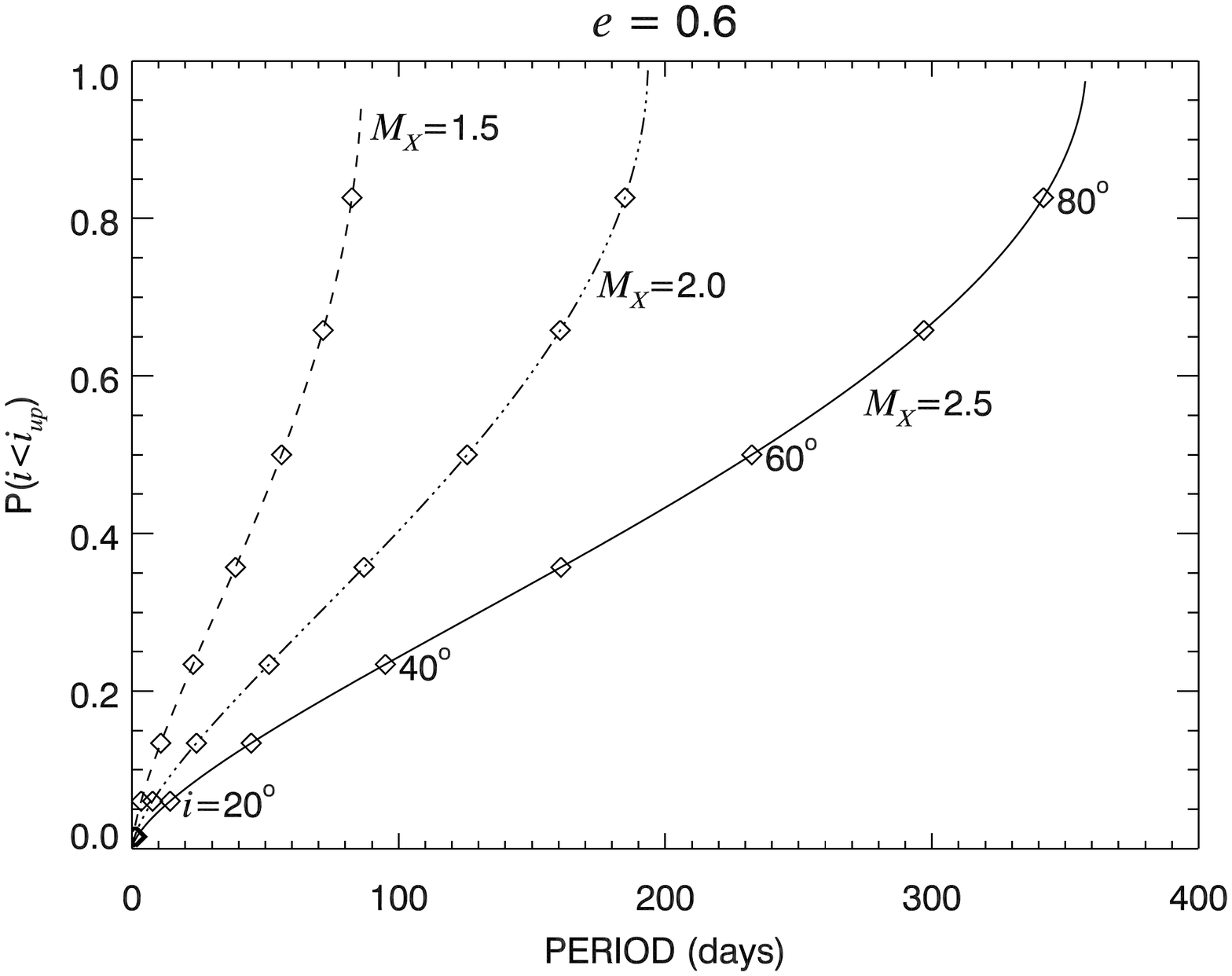}
\caption{Probability to have missed the binary companion because of a too low inclination angle as a function of the orbital period, for eccentricities equal to 0.0, 0.2, 0.4, and 0.6, respectively. In each panel, three different secondary masses have been assumed.  The diamonds indicate inclinations from 10$^{\circ}$ to 80$^{\circ}$, spaced at intervals of 10$^{\circ}$.
\label{constraints}  }
\end{figure}

We used the lower $\sigma_{RV}$ obtained from the OHP spectra and our measured mass of the Be star, $M_\star = 16 \: M_\odot$ (see Section 5).  Secondary masses of 1.5, 2.0, and 2.5 M$_{\odot}$, consistent with a low mass companion, and eccentricities of 0.0, 0.2, 0.4, and 0.6 were considered.  The results are plotted in Figure \ref{constraints}.  
For a compact companion of 1.5 M$_\odot$, we would be unable to detect $V_R$ shifts for an orbital period greater than 50 days.  The chance of detection improves for higher values of the secondary mass.  However, for a 100 day orbit, there is still only a 50\% probability the the system will have an inclination favorable for detecting $V_R$ shifts in our data.  The chances of detection also decrease with orbital eccentricity.  However, our method does not take into account that to detect the radial velocity shifts for an eccentric orbit, we would need to observe the system close to periastron.  The system would only spend a small fraction of its orbit with $V_R$ shifts high enough for us to detect, decreasing our chances of observing $V_R$ shifts in systems of higher eccentricity.  Therefore, we cannot rule out HD 259440 as a binary system with a long period.


\section{Circumstellar Disk Features \label{df}}

\begin{figure}
\includegraphics[angle=0,scale=0.50]{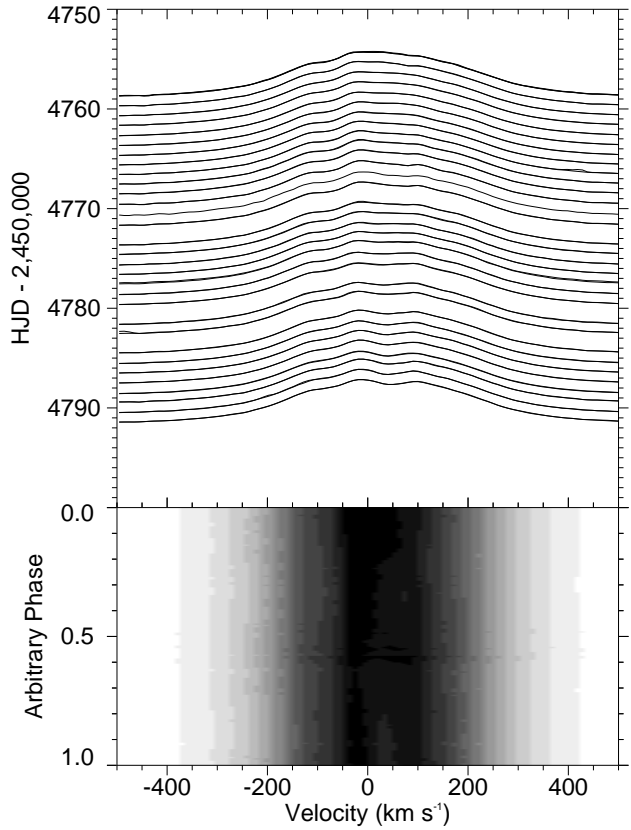}
\\
\caption{The upper plot shows the H$\alpha$ line profile of HD 259440 over our continuous 35 nights of observation with the CF, sorted by HJD, and the lower plot shows a gray-scale image of the same line.  Since no period is currently known, the ``Arbitrary Phase" goes from 0 at the start of the observing run to 1.0 at the end of the run. The intensity at each velocity in the gray-scale image is assigned one of 16 gray levels based on its value between the minimum (bright) and maximum (dark) observed values.  The intensity between observed spectra is calculated by a linear interpolation between the closest observed phases.
\label{N1} }
\end{figure}

Our collection of red CF spectra provide an excellent tracer of the Be star's disk emission over 35 nights.  The mean equivalent width of the H$\alpha$ emission line, $W_{\rm H\alpha}$, is $-52.3$ \AA, with a standard deviation of 0.7.  We find that $W_{\rm H\alpha}$ is constant within our estimated 2\% error due to noise and the continuum placement.  We show in Figure \ref{N1} the H$\alpha$ line profiles and a gray-scale image of this line over the span of the observing run.  Since the orbital period of HD 259440 is unknown, neither the line profiles nor the gray-scale plots are folded by orbital phase, but rather they reveal true chronological variations in the line profile behavior as a function of heliocentric julian date (HJD).

\begin{figure}
\includegraphics[angle=0,scale=0.50]{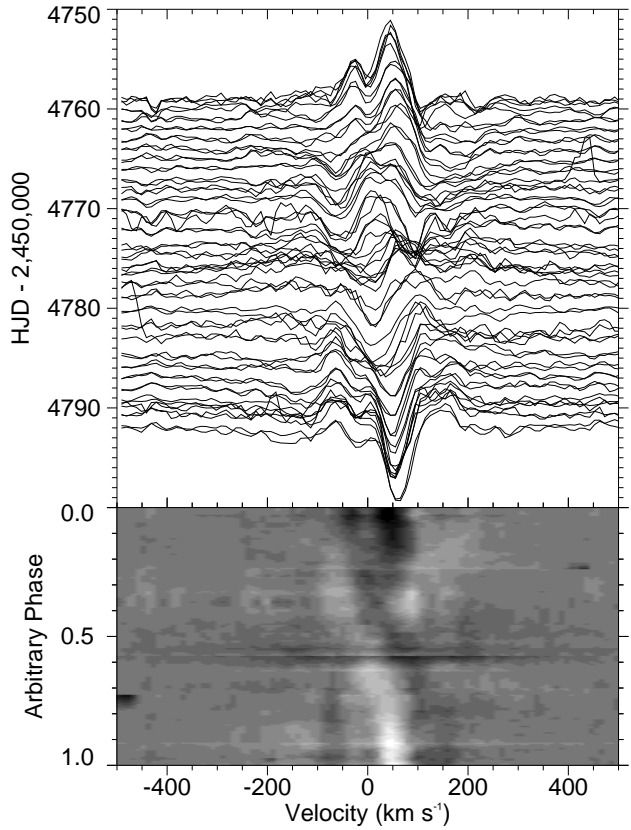}
\\
\caption{The upper panel shows the H$\alpha$ residual spectra of HD 259440, and the bottom panel shows a grayscale plot of the same residuals, in the same format as Fig.\ \ref{N1}.  The residual spectra reveal  variations in the H$\alpha$ line with a $\sim 60$ d period that we attribute to a spiral density wave in the circumstellar disk.
\label{N2} }
\end{figure}

We noticed some subtle temporal variations in the shape of the emission profile, so we subtracted the mean emission line profile to investigate the residuals more carefully.  The H$\alpha$ emission residuals, and the corresponding gray-scales, are shown in Figure \ref{N2}.  These residual spectra reveal a partial S-shaped feature over the 35 nights of observation that suggest a characteristic period of $\sim 60$ days.  We attribute these variations to a spiral density wave in the circumstellar disk, common among Be stars \citep{porter}.

The high $V \sin i$ for this star (see Section 5) implies that the disk inclination must be nearly $90^\circ$.  Therefore we view the disk nearly edge-on.  Since we do not resolve a double peaked profile in H$\alpha$, the disk is likely optically thick out to several stellar radii (consistent with the very high $W_{\rm H\alpha}$).  By contrast, H$\gamma$ shows a strong double-peaked emission profile.  The disk also contributes a continuum flux to the line of sight which is added to the observed stellar spectrum, causing photospheric lines in the rectified spectra to appear too shallow.  Bound-free emission will contribute a fraction of the stellar flux across all wavelengths, while free-free emission will increase at longer wavelengths.  Indeed, we find evidence for such flux dilution in our blue spectra, discussed in the following section.  We also observe an infrared excess in the SED for the system, which will be discussed in Section 6.


\section{Stellar Parameters \label{stelpar}}

\begin{figure}
\includegraphics[angle=180,scale=0.45]{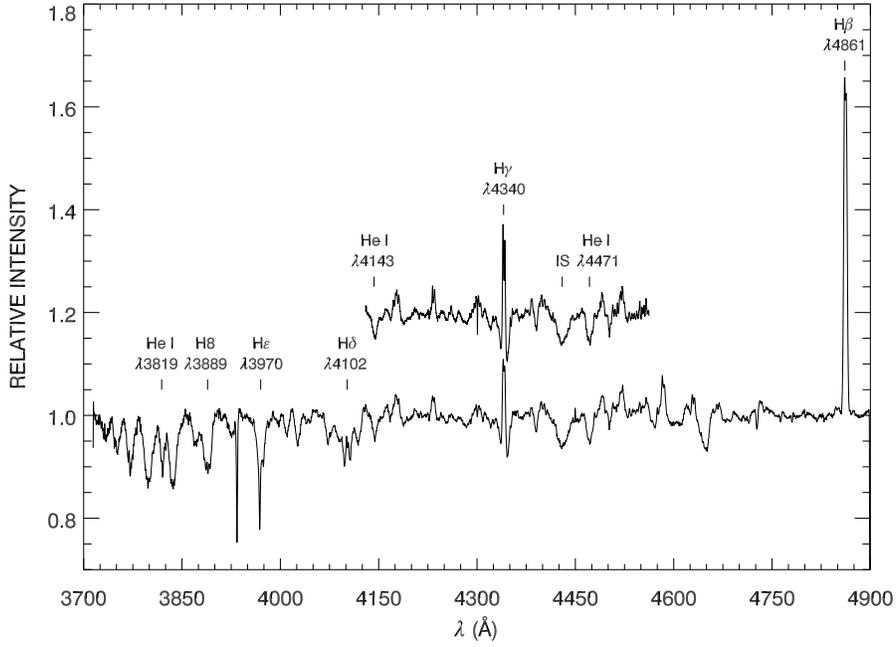} 
\\
\caption{The mean spectra from the KPNO 2.1m (bottom) and CF (top) telescopes are plotted for comparison.  The CF spectrum is offset by 0.2 for clarity.
\label{KPNOplot} }
\end{figure}

Accurately determining the stellar parameters of HD 259440 is a challenging process due to the prevalence of emission features in the spectra.  In order to minimize the impact of the disk emission on our analysis, we chose to focus on the \ion{He}{1} and H Balmer lines with $\lambda \le$ 4000 \AA, where the lines are predominantly in absorption.   This approach required our analysis to use the lower resolution KPNO 2.1m telescope spectra, as these covered a broader wavelength range than the KPNO CF data (Figure \ref{KPNOplot}).  In order to improve the S/N, we created a mean spectrum from the two observations.  We then compared this spectrum to Tlusty OSTAR2002 \citep{lanz2003} and BSTAR2006 \citep{lanz2007} model grids, corrected for instrumental broadening, to obtain values for the effective temperature, $T_{\rm eff}$, surface gravity, $\log g$, and $V \sin i$ of the optical star.  

The Tlusty BSTAR2006 and OSTAR2002 model grids assume a plane-parallel stellar atmosphere.  The BSTAR2006 models are available with temperatures ranging from 15000 K to 30000 K, with a grid spacing of 1000 K, and $\log g$ ranging from 1.75 to 4.75 where g is in cm s$^{-2}$, with a grid spacing of 0.25.  The OSTAR2002 models exist for temperatures between 27500 K and 55000 K with a grid spacing of 2500 K.  These models are available for $\log g$ between 3.0 and 4.75, with a grid spacing of 0.25.  The OSTAR2002 models assume a mictroturbulence velocity of 10 km s$^{-1}$, while the BSTAR2006 models assume a microturbulence velocity of 2 km s$^{-1}$.  We used models with solar abundances.

Initially, we tried fitting the mean spectrum of HD 259440 over the 3750$-$3950 \AA~range for $T_{\rm eff}$, $\log g$, and $V \sin i$ simultaneously.  $T_{\rm eff}$ was varied between 20000 K and 40000 K in increments of 1000 K (2500 K) to match the spacing of the BSTAR2006 (OSTAR2002) grid.  We varied $\log g$ from 2.50 to 4.75 (3.0 to 4.75) in increments of 0.25 according to the BSTAR2006 (OSTAR2002) grid spacings.  We also increased $V \sin i$ from 300 km s$^{-1}$ to 500 km s$^{-1}$ in steps of 50 km s$^{-1}$.  We computed the $\chi^{2}$ value for each combination of $T_{\rm eff}$, log $g$, and $V \sin i$.   However, none of the trials produced a satisfactory fit to the observed spectrum.  The best fits from the two model grids produced divergent solutions for $T_{\rm eff}$ and $\log g$.  Furthermore, a visual inspection comparing the model to the observed spectrum revealed the poor quality of these fits and they could not be used to constrain a $T_{\rm eff}$ and $\log g$ for the system.

The unusually shallow yet broad absorption lines seen in the spectra of HD 259440 are indicative of flux dilution, in that the continuum emission and scattering from the disk has effectively decreased the relative strength of absorption lines in the rectified spectra.  The total observed flux is the sum of the flux from the disk, $F_{disk}$, and the star, $F_{star}$.  Over a small wavelength range, we can assume a constant ratio $N=F_{disk}$/$F_{star}$.  Using this assumption, we can correct for the disk continuum emission by adjusting the rectified spectrum:
\begin{equation}
F_{R,star} = F_{R} (1+N) - N
\end{equation}
where $F_{R}$ is the rectified observed spectrum and $F_{R,star}$ is the rectified spectrum of the star corrected for the disk continuum emission.  Since the relative disk flux is unknown, we adjusted the observed spectrum for values of $N = 0.1-0.9$ and repeated the fitting procedure at each value.

The absorption lines in the spectrum of HD 259440 are fairly broad, so we expect the star to have a high rotational velocity.  At all values of $T_{\rm eff}$ and $\log g$, a $V \sin i$ of 500 km s$^{-1}$ provided the best agreement with the line profiles.  Therefore, we chose to fix $V \sin i$ at this value.  We then repeated our fits using the OSTAR2002 and BSTAR2006 models over the 3750$-$3950 \AA \ region.  At each $N$, we determined the best fit $T_{\rm eff}$ and $\log g$ by minimizing the $\chi^{2}$ from the model fits.  Our quoted errors for $T_{\rm eff}$ and $\log g$ represent the $1\sigma$ significance levels from our best fit.   For values of $N$ between 0.0 and 0.4, $T_{\rm eff}$ and $\log g$ are not well constrained, but the fit quality and agreement between the BSTAR2006 and OSTAR2002 models improve significantly for higher values of $N$.  Therefore, we chose to focus on the results with $N \geq 0.5$, which are summarized in columns 1$-$5 of Table \ref{stellarparams}.  

\begin{figure}
\includegraphics[angle=180,scale=0.45]{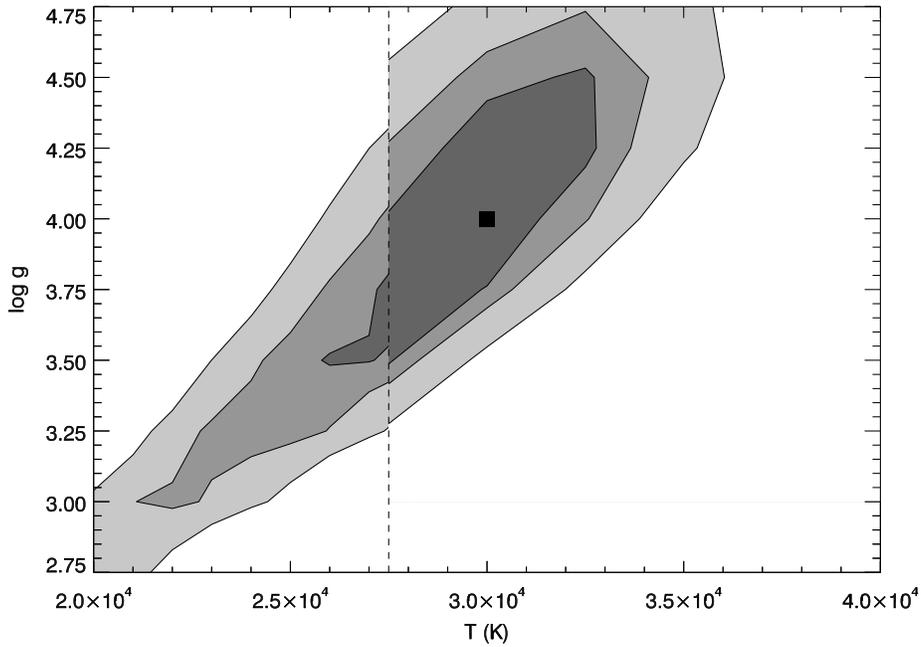} 
\\
\caption{A contour plot of the $1\sigma$, $2\sigma$, and $3\sigma$ errors from the spectral fits using $N$ = 0.7 and $V \sin i$ = 500 km s$^{-1}$.  The discontinuity between the OSTAR2002 and BSTAR2006 models is marked by the vertical dashed line.  The square represents the model with the lowest $\chi^2$ value for both model fits.
\label{contourOB7} }
\end{figure}

A contour plot illustrating the $1\sigma$, $2\sigma$, and $3\sigma$ significance errors for the combined OSTAR2002 and BSTAR2006 model fits in the $N=0.7$ case is shown in Figure \ref{contourOB7}.  As can be seen from this figure and Table \ref{stellarparams}, the best values of $T_{\rm eff}$ for both the OSTAR2002 and BSTAR2006 models lie near the boundaries of the model grids.  Furthermore, the $1\sigma$ significance contours are not closed at the boundary between the models.  Therefore, we determined the lower limit of $T_{\rm eff}$ from the $1\sigma$ contour of the BSTAR2006 models and the upper limit from the $1\sigma$ contour of the OSTAR2002 models.  

\begin{figure}
\includegraphics[angle=180,scale=0.50]{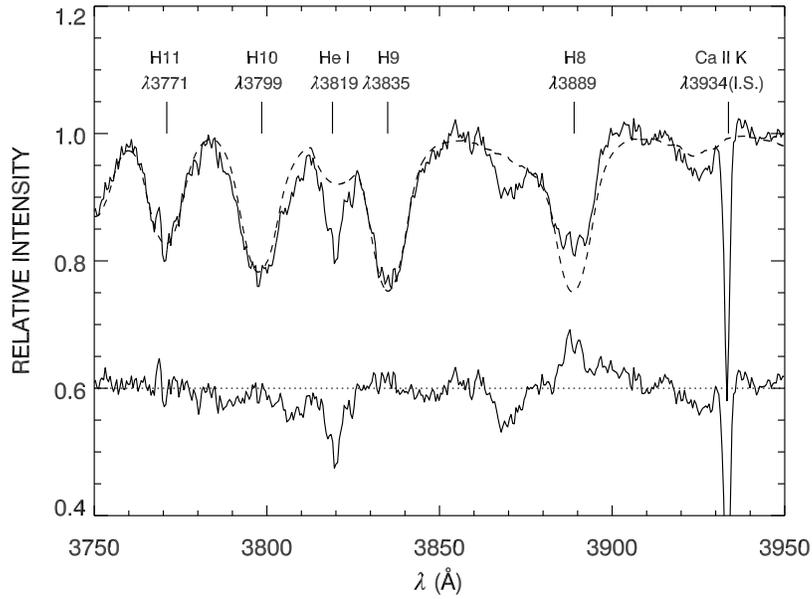} 
\\
\caption{The mean KPNO 2.1m spectrum, adjusted for $N=0.7$, is shown as a solid line.  The dashed line shows the Tlusty BSTAR2006 model spectrum with $T_{\rm eff}$ = 30000 K, log $g$ = 4.0, and $V \sin i$ = 500 km s$^{-1}$.  The residuals from the fit are plotted below the spectrum, shifted by 0.6 for easier comparison.  
\label{model7} }
\end{figure}

A Tlusty BSTAR2006 model spectrum with $T_{\rm eff}$ = 30000 K, log $g$ = 4.0, and $V \sin i$ = 500 km s$^{-1}$ plotted against the mean spectrum of HD 259440 with $N=0.7$ is shown in Figure \ref{model7}.  We generally find good agreement between the model and observed spectrum in the H Balmer lines, although the \ion{He}{1} $\lambda 3819$ line appears stronger than all of our  model fits.  HD 259440 has been classified as a star with peculiar abundances.  It was marked as nitrogen rich by \citet{turner} and may also be a helium-strong star.  We cannot distinguish significant differences between the model fits with different values of $N$, hence some ambiguity in the physical parameters of this system.  

\begin{figure}
\includegraphics[angle=180,scale=0.50]{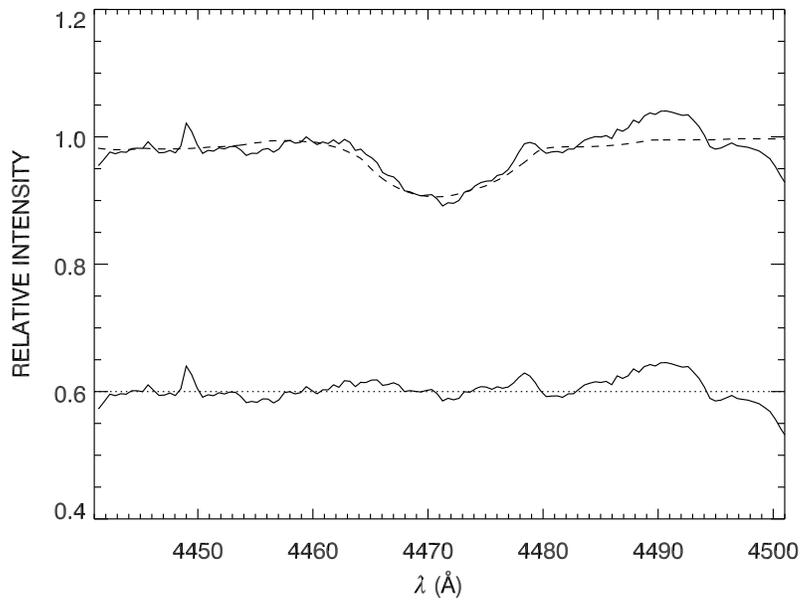} 
\\
\caption{The \ion{He}{1} $\lambda 4471$ line from the mean KPNO 2.1m spectrum is shown in the same format as Figure \ref{model7}. 
\label{model7He} }
\end{figure}

As a further test of the validity of the fit, we compared the model fits to one of the few other photospheric absorption lines in our blue spectra, the flux-corrected \ion{He}{1} $\lambda 4471$ line. Figure \ref{model7He} shows good agreement over the core of the line with the BSTAR2006 fit and $N = 0.7$ .  There may be some excess emission in the line wings from the higher velocity regions of the disk, narrowing the line slightly.  Using the OSTAR2002 fits, the flux-corrected \ion{He}{1} $\lambda 4471$ line is again stronger than the model, a result of the difference in microturbulent velocity between the two sets of models.  Since HD 259440 may be He strong, the fit of the \ion{He}{1} $\lambda 4471$ line does not help constrain the best fit any further.  The quality of the fits for the values of $N$ considered in this paper were equally good, so we also cannot constrain the flux contribution from the disk any further.

We determined the mass, $M_{\star}$, and radius, $R_{\star}$, of the star using our best fit values for $T_{\rm eff}$ and log $g$ and interpolating between the evolutionary tracks of \citet{schaller}.  The evolutionary tracks were published for non-rotating stars with solar metallicity and masses ranging from 0.8 to 120 M$_{\odot}$.  The errors for  $M_{\star}$ were calculated using the best value of $\log g$ and using the $1 \sigma$ limits for $T_{\rm eff}$, while the errors for $R_{\star}$ were calculated using the best fit values for $T_{\rm eff}$ and using the $1 \sigma$ limits for $\log g$.  Our measurements of $M_\star$ and $R_\star$ are presented in columns 6$-$9 of Table \ref{stellarparams}.  The rapid rotation of HD 259440 may result in an underestimation of the mass and introduce an additional systematic error into our measurements.  \citet{ekstrom} calculated evolutionary tracks for rapidly rotating stars.  Comparing these models to the tracks from \citet{schaller}, we conclude that any discrepancy in the mass of HD 259440 would be about 5 M$_{\odot}$.  We also calculated the critical velocity, $V_{crit}$, at the stellar equator for each $N$.  Since HD 259440 is a rapid rotator, the stellar radius at the equator is $1.5R_\star$ in the standard Roche model.  Therefore,
\begin{equation}
V_{crit}=\sqrt{\frac{GM_\star}{1.5R_\star}}.
\end{equation}
The results are listed in column 10 of Table \ref{stellarparams}.  These results indicate that HD 259440 is rotating close to $V_{crit}$, which is plausible for Be stars and thought to contribute to the formation of the circumstellar disk \citep{townsend}.


\section{Spectral Energy Distribution \label{sed}}

We compared the observed SED of HD 259440 with the model SEDs from the Tlusty BSTAR2006 and OSTAR2002 models to measure the distance to the star.  We obtained ultraviolet broadband photometric fluxes from \citet{wesselius}, Johnson $UBV$ magnitudes from \citet{neckel}, Johnson $I$ magnitude from \citet{droege}, 2MASS $JHK_S$ magnitudes from \citet{cutri}, and mid-infrared fluxes from \citet{egan}.  The Johnson magnitudes were converted to fluxes according to \citet{bessel}, while the 2MASS photometry were converted to fluxes using \citet{cohen}.

The Tlusty model spectra were binned in 50 \AA \ bins to eliminate small scale line features and simplify comparison with the observed fluxes.  A reddening of $E(B-V) = 0.85$ \citep{friedemann} and selective extinction of $R = 3.1$ were applied to the model SEDs using the extinction model of \citet{fitzpatrick}.   The observed photometric fluxes were corrected for disk emission using each value of $N$ and compared with the best fit OSTAR2002 and BSTAR2006 models. 
 
In our spectral fitting, we assumed that the disk flux ratio $N$ is constant over a small wavelength range.  Therefore, we compared the observed fluxes between $3000-5000$ \AA, corrected for $N$, to the model SEDs to determine the angular diameter, $\theta_R$:
\begin{equation}
\theta_R = \frac{R_\star}{d} = \sqrt{\frac{F_{\star, obs}}{F_{\star, surf}}} = \sqrt{\frac{F_{tot,obs}}{F_{\star, surf}(1+N)}}
\end{equation}
where $F_{\star,obs}$ is the observed flux produced by the star, $F_{\star, surf}$ is the flux at the stellar surface, $F_{tot, obs}$ is the combined flux of the star and disk, and $d$ is the distance to the star.  Using our derived $R_\star$ with the calculated $\theta_R$, we find that HD 259440 lies within the distance range $d = 1.1-1.7$ kpc.   The errors in the distance presented in Table \ref{stellarparams} give the standard deviation between the model SED and the broadband photometric flux points in the $3000-5000$ \AA~range.  As the quality of all our best fits were similar for values of $N$ between 0.5 and 0.9, the actual error in $d$ is more accurately reflected by the range of values presented.  HD 259440 is listed by \citet{voroshilov} as a member of the open cluster NGC 2244.  A recent study by \citet{bonatto} places the cluster at 1.6 $\pm$ 0.2 kpc from the Sun, a value consistent with previous studies and our measurements for HD 259440 \citep{ogura,perez,park}.

\begin{figure}
\includegraphics[angle=180,scale=0.50]{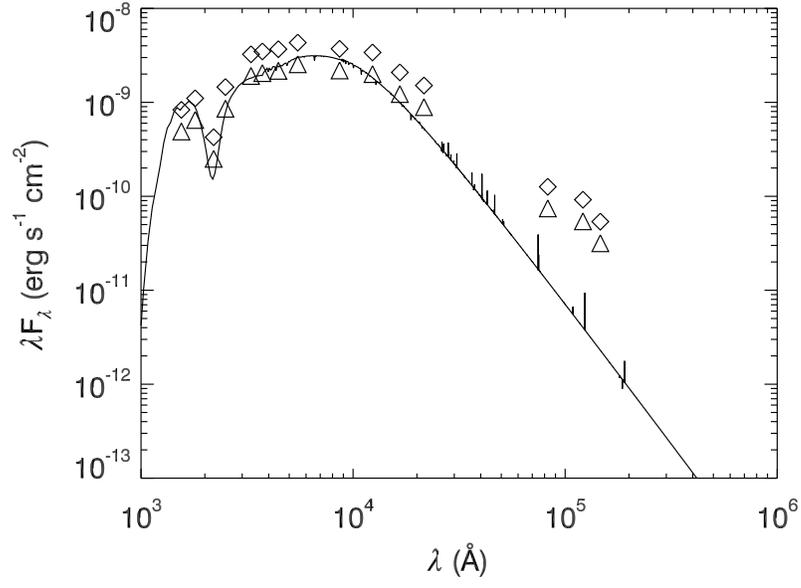} 
\\
\caption{The Tlusty BSTAR2006 model SED normalized to $\theta_R$ = 24.3 $\mu$as is plotted as a solid line.  The diamonds indicate the observed broadband photometric fluxes.  The triangles represent the estimated stellar flux assuming $N=0.7$.
\label{sedB7B} }
\end{figure}

We expect to see exorbitant infrared emission in the SED of HD 259440 due to the nearly edge-on circumstellar disk.  This infrared excess can clearly be seen in Figure \ref{sedB7B}, which shows the observed broadband photometric fluxes with the BSTAR2006 model SED for $T_{\rm eff} = 30000$ K, $\log g= 4.0$, $N=0.7$, and $d = 1.30$ kpc.  At blue optical wavelengths, the disk contributes $\sim 0.7 F_\star$.  In the mid-infrared, the disk continuum emission is $\sim 22$ times the flux of the star.


\section{Summary of Results \label{conclusions}}

HD 259440 is a main sequence B0pe star that may be helium strong.  Due to continuum emission and scattered light from the disk contaminating the stellar spectrum, our measurements of its physical parameters are somewhat ambiguous.  We find $T_{\rm eff} \approx 30000$ K and $\log g \approx 4.0$.  From the evolutionary tracks of Schaller et al. (1992), our best model fits indicate that the star has a mass of about 16 $M_\odot$ and radius of about 6.6 $R_\odot$.  We find an angular radius of 23.5$-$25.7 $\mu$as and a distance range of 1.1$-$1.7 kpc.  Due to the non-detection of radial velocity shifts over our 35 nights of CF data, we can rule out the possible 35$-$40 d orbital period suggested by Falcone et al. (2010).  However, longer period orbits would produce $V_R$ shifts that would be lost in the noise of our data sets.  Therefore, we cannot rule out HD 259440 as a binary with a period greater than 100 days, which would be consistent with the class of non-accreting X-ray binaries.

\acknowledgments
We would like to thank Wenjin Huang for allowing us to obtain spectra of HD 259440 during his KPNO 2.1m telescope time in December 2008.  We also thank Yves Gallant, Bertrand Plezv, and Guillaume Dubus, whose observations comprise the OHP archival spectra used in this paper.  Also, we appreciate helpful conversations with Marc Gagn\'e, Stan Owocki, Guillame Dubus, and Doug Gies about this work.  We are grateful for support from NASA DPR numbers NNX08AV70G, NNG08E1671, and NNX09AT67G.  MVM is grateful for an institutional grant from Lehigh University.  This publication makes use of data products from the Two Micron All Sky Survey, which is a joint project of the University of Massachusetts and the Infrared Processing and Analysis Center/California Institute of Technology, funded by the National Aeronautics and Space Administration and the National Science Foundation.



{\it Facilities:} \facility{KPNO:2.1m ()}, \facility{KPNO:CFT ()}, \facility{OHP:1.93m ()}

\clearpage


\clearpage

\begin{deluxetable}{lcccccc}
\small
\tablewidth{0pt}
\tablecaption{Journal of Spectroscopy \label{obs}}
\tablehead{
\colhead{UT Dates} &
\colhead{Telescope} &
\colhead{$\lambda$ Range} &
\colhead{Resolving Power} &
\colhead{Grating/} &
\colhead{Filter} &
\colhead{Number of} \\
\colhead{} &
\colhead{} &
\colhead{(\AA)} &
\colhead{$(\lambda/\Delta \lambda)$} &
\colhead{Order} &
\colhead{} &
\colhead{Spectra} }
\startdata
2007 Oct$-$2008 Feb 	& OHP 1.93m 	& $3872-6943$ 	& 40000 			& R2/50--88 		& \nodata 		& 16 \\
2008 Oct 17$-$Nov 21 	& KPNO CF 	& $4130-4570$ 	& 9500 			& B/3 			& 4-96 		& 68 \\
2008 Oct 17$-$Nov 21 	& KPNO CF 	& $6400-7050$ 	& 12000 			& B/2 			& OG550 		& 62 \\
2008 Dec 12$-$13 		& KPNO 2.1m 	& $3700-4900$ 	& 2100$-$3100 	& G47/2 			& CuSO$_{4}$ & 2 \\
2009 Mar 				& OHP 1.93m 	& $3872-6943$ 	& 40000 			& R2/50--88 		& \nodata		& 2 \\
2009 Oct 				& OHP 1.93m 	& $3872-6943$ 	& 40000 			& R2/50--88 		& \nodata 		& 5 
\enddata
\end{deluxetable}

\begin{deluxetable}{lcc}
\tablewidth{0pt}
\tablecaption{OHP $V_R$ Measurements \label{vrmes}}
\tablehead{
\colhead{UT Date} &
\colhead{Mid-exposure} &
\colhead{$V_R$} \\
\colhead{} &
\colhead{HJD$-2450000$} &
\colhead{km s$^{-1}$} }
\startdata
2007 Oct 02 & 4375.1714 & 43.78 \\
2007 Oct 22 & 4395.1686 & 47.55 \\
2007 Oct 22 & 4395.1744 & 46.69 \\
2007 Nov 01 &	4405.1763 & 47.96 \\
2007 Nov 02 &	4406.1672 & 46.48 \\
2007 Nov 03 &	4407.1905 & 44.90 \\
2007 Nov 04 &	4408.1886 & 47.04 \\
2007 Nov 05 &	4409.1474 & 41.74 \\
2007 Nov 10 &	4414.0348 & 47.81 \\
2007 Dec 27 &	4461.9402 & 41.99 \\
2007 Jan 19 & 4484.8059 & 38.06 \\
2008 Jan 21 & 4486.8267 & 28.88 \\
2008 Feb 06 & 4502.9462 & 30.41 \\
2008 Feb 12 & 4508.8925 & 36.28 \\
2008 Feb 19 & 4515.8693 & 31.02 \\
2008 Feb 25 & 4521.8527 & 27.09 \\
2009 Mar 10 & 4900.8138 & 33.11 \\
2009 Mar 14 & 4904.7721 & 45.97 \\
2009 Oct 13 & 5117.6426 & 39.85 \\
2009 Oct 15 & 5119.6701 & 43.57 \\
2009 Oct 17 & 5121.6604 & 41.69 \\
2009 Oct 27 & 5131.6605 & 42.96 \\ 
2009 Oct 28 & 5132.6867 & 47.04 \\
\enddata
\end{deluxetable}


\begin{deluxetable}{lccccccccccccc}
\tablecolumns{14}
\tablewidth{0pt}
\tablecaption{Stellar Parameters \label{stellarparams}} 
\tablehead{
\colhead{$N$} & 
\colhead{$T_{\rm eff}$} & 
\colhead{$\Delta T_{\rm eff}$} & 
\colhead{$\log g$} & 
\colhead{$\Delta \log g$} &
\colhead{$M_{\star}$} &                                                                                       
\colhead{$\Delta M_{\star}$} &
\colhead{$R_{\star}$} &
\colhead{$\Delta R_{\star}$} &
\colhead{$V_{crit}$} &
\colhead{$\theta_{D}$} &
\colhead{$\Delta \theta_{D}$} &
\colhead{$d$} &
\colhead{$\Delta d$} \\
\colhead{} & 
\colhead{(K)} & 
\colhead{(K)} & 
\colhead{($dex$)} & 
\colhead{($dex$)} &
\colhead{(M$_{\odot}$)} &
\colhead{(M$_{\odot}$)} &
\colhead{(R$_{\odot}$)} &
\colhead{(R$_{\odot}$)} &
\colhead{(km s$^{-1}$)} &
\colhead{($\mu$as)} &
\colhead{($\mu$as)} &
\colhead{(kpc)} &
\colhead{(kpc)} }
\startdata
\sidehead{BSTAR2006 Model Fitting Results}
0.5 & 30,000 & $_{-7,000}^{+1,000}$ & 3.75  & $_{-0.85}^{+0.90}$ & 19.0 & $_{-8.2}^{+1.5}$ & 9.6 & $_{-6.9}^{+44.8}$ & 501 & 25.1 & $\pm 1.1$ & 1.78 & $\pm 0.08$ \\
0.6 & 29,000 & $_{-3,000}^{+4,000}$ & 3.75 & $_{-0.30}^{+0.80}$ & 17.6 & $_{-4.0}^{+6.7}$ & 9.3 & $_{-6.4}^{+5.9}$ & 491 & 25.2 & $\pm 1.2$ & 1.71 & $\pm 0.09$ \\
0.7 & 30,000 & $_{-5,000}^{+3,000}$ & 4.00 & $_{-0.50}^{+0.50}$ & 16.0 & $_{-5.0}^{+3.6}$ & 6.6 & $_{-3.3}^{+8.1}$ & 555 & 23.7 & $\pm 1.1$ & 1.30 & $\pm 0.06$ \\
0.8 & 30,000 & $_{-5,000}^{+2,000}$ & 4.00 & $_{-0.55}^{+0.50}$ & 16.0 & $_{-5.0}^{+2.5}$ & 6.6 & $_{-3.3}^{+9.7}$ & 555 & 23.1 & $\pm 1.1 $ & 1.33 & $\pm 0.06$ \\
0.9 & 29,000 & $_{-6,000}^{+2,000}$ & 4.00 & $_{-0.60}^{+0.45}$ & 14.5 & $_{-5.3}^{+2.9}$ & 6.3 & $_{-2.9}^{+7.7}$ & 541 & 23.4 & $\pm 1.2$ & 1.26 & $\pm 0.07$ \\
\sidehead{OSTAR2002 Model Fitting Results}
0.5 & 30,000 & $_{-7,000}^{+1,000}$ & 4.00  & $_{-1.10}^{+0.65}$ & 16.0 & $_{-6.8}^{+1.4}$ & 6.6 & $_{-3.9}^{+47.8}$ & 555 & 24.5 & $\pm 1.0$ & 1.25 & $\pm 0.05$ \\
0.6 & 30,000 & $_{-4,000}^{+3,000}$ & 4.00 & $_{-0.55}^{+0.55}$ & 16.0 & $_{-4.4}^{+3.6}$ & 6.6 & $_{-3.5}^{+9.7}$ & 555 & 23.7 & $\pm 0.9$ & 1.30 & $\pm 0.05$ \\
0.7 & 30,000 & $_{-5,000}^{+3,000}$ & 4.00 & $_{-0.50}^{+0.50}$ & 16.0 & $_{-5.0}^{+3.6}$ & 6.6 & $_{-3.3}^{+8.1}$ & 555 & 23.0 & $\pm 0.9$ & 1.34 & $\pm 0.05$ \\
0.8 & 27,500 & $_{-2,500}^{+4,500}$ & 3.75 & $_{-0.30}^{+0.60}$ & 15.3 & $_{-2.8}^{+7.2}$ & 8.6 & $_{-5.6}^{+5.5}$ & 476 & 24.5 & $\pm 1.2$ & 1.64 & $\pm 0.08$ \\
0.9 & 27,500 & $_{-4,500}^{+3,500}$ & 4.00 & $_{-0.60}^{+0.45}$ & 13.2 & $_{-4.0}^{+4.2}$ & 6.0 & $_{-2.8}^{+9.4}$ & 529 & 24.1 & $\pm 1.3$ & 1.16 & $\pm 0.06$ \\
\enddata
\tablecomments{The errors in $T_{\rm eff}$ and $\log g$ were calculated using the OSTAR2002 results to determine the upper limits and the BSTAR2006 results to determine the lower limits.}
\end{deluxetable}

\end{document}